\documentclass[%
 aip, apl,%
 sd,%
 amsmath,amssymb,
 reprint,%
]{revtex4-1}

\usepackage{graphicx}
\usepackage{dcolumn}
\usepackage{bm}
\usepackage{ulem}
\usepackage{tabu}
\usepackage{natmove}

\usepackage{color}

\begin{document}

\title{Charged Vacancy Defects in Black Phosphorus Monolayer Phosphorene}
\author{Biswas Rijal}
\email{biswas.rijal@ufl.edu}
\affiliation{Materials Science and Engineering, University of Florida, Gainesville, Florida 32611, U.S.A}
\affiliation{Quantum Theory Project, University of Florida, Gainesville, Florida 32611, U.S.A}
\author{Anne Marie Z. Tan}
\affiliation{School of Mechanical and Aerospace Engineering, Nanyang Technological University, Singapore 639798, Singapore}
\affiliation{Materials Science and Engineering, University of Florida, Gainesville, Florida 32611, U.S.A}
\affiliation{Quantum Theory Project, University of Florida, Gainesville, Florida 32611, U.S.A}
\author{Christoph Freysoldt}
\affiliation{Max Planck Institut f\"ur Eisenforschung, D\"usseldorf, Germany}
\author{Richard G. Hennig}
\email{rhennig@ufl.edu}
\affiliation{Materials Science and Engineering, University of Florida, Gainesville, Florida 32611, U.S.A}
\affiliation{Quantum Theory Project, University of Florida, Gainesville, Florida 32611, U.S.A}
\date{\today}

\begin{abstract}
  The two-dimensional semiconductor phosphorene has attracted extensive research interests for potential applications in optoelectronics, spintronics, catalysis, sensors, and energy conversion. To harness phosphorene's potential requires a better understanding of how intrinsic defects control carrier concentration, character, and mobility. Using density-functional theory and a charge correction scheme to account for the appropriate boundary conditions, we conduct a comprehensive study of the effect of structure on the formation energy, electronic structure and charge transition level of the charged vacancy point defects in phosphorene. We predict that the neutral vacancy exhibits a 9-5 ring structure with a formation energy of 1.7~eV and transitions to a negatively charged state at a Fermi level 1.04~eV above the valence band maximum. The corresponding optical charge transitions display sizeable Frank-Condon shifts with a large Stokes shift of 0.3~eV. Phosphorene vacancies should become negatively charged in $n$-doped phosphorene, which would passivate the dopants and reduce the charge carrier concentration and mobility.
\end{abstract}
\maketitle

\section{Introduction}

\begin{figure}[t]
    \includegraphics[width=8cm]{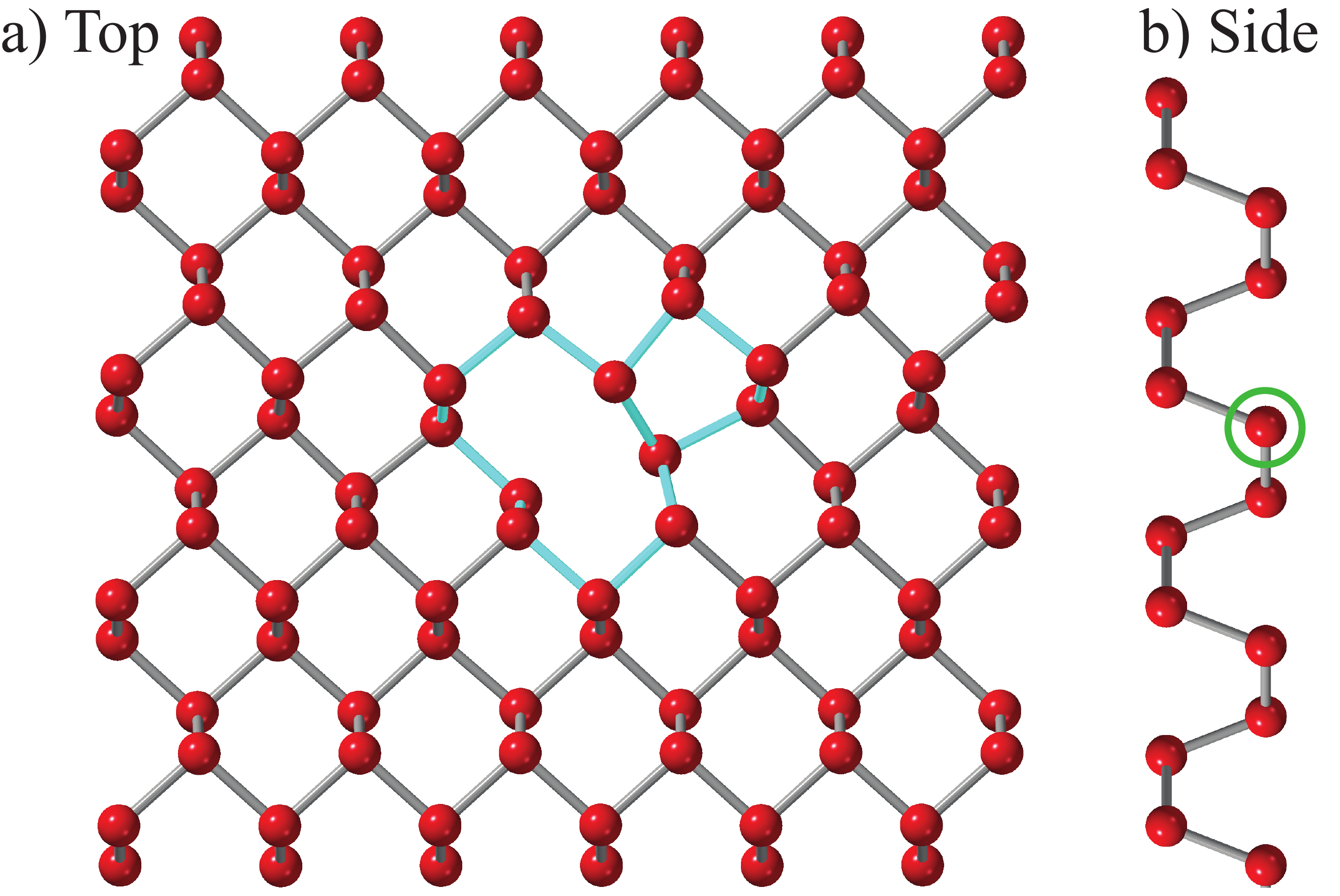}
    \caption{\label{fig:phosphorene}(Color online) Structure of 2D phosphorene with a single vacancy point defect. a) The top view shows the 9-5 ring (turquoise lines) formed by relaxation of the defect structure. b) The side view shows the unrelaxed structure, and the green circle indicates the atom that is removed to create the vacancy.} 
\end {figure}

Black phosphorus is the most stable allotrope of phosphorus and was synthesized accidentally in 1914 when white phosphorus was subjected to the high pressure of 1.2~GPa and a temperature of 473~K~\cite{bridgman1914two, ribeiro2018raman, ling2015renaissance}. Like graphite, black phosphorus is a layered crystal held together by weak van der Waals forces. One hundred years after its discovery, black phosphorus was mechanically exfoliated into a few-layered form~\cite{li2014black,liu2014phosphorene}. This two-dimensional (2D) form of phosphorus, named phosphorene, is the only elemental 2D material besides graphene that can be mechanically exfoliated~\cite{liu2014phosphorene}. Phosphorene has been widely studied for various applications in energy technology~\cite{li2020recent,batmunkh2018black}, electronic devices~\cite{liu2014phosphorene}, catalysis~\cite{ran2018metal,wang2016quantum}, and sensors~\cite{liu2017two,buscema2014fast}, due to its high carrier mobility~\cite{liu2014phosphorene}, optical and electronic anisotropy~\cite{tran2014layer,xia2014rediscovering}, and tunable bandgap~\cite{tran2014layer}. Fig.~\ref{fig:phosphorene} shows that single-layer phosphorene displays an anisotropic puckered honeycomb structure comprised of two layers of phosphorus atoms bonded by $sp^3$ hybridized orbitals~\cite{rudenko2014quasiparticle,boulfelfel2012squeezing}.

For the use of phosphorene in electronic devices, it is paramount to determine the formation energy of point defects and understand their effect on the electronic properties. Obtaining this information from experiments alone is difficult due to the high concentration of defects and impurities introduced during synthesis and the difficulty in establishing thermodynamic equilibrium. Computational methods, such as density-functional theory (DFT), provide a reliable approach to predict defect properties in semiconductors and complement experiments~\cite{van2004first, freysoldt2014first, komsa2014charged, freysoldt2018first}. For DFT calculations of charged defects in 2D materials with plane-wave basis sets, unphysical electric fields appear in the vacuum region of supercell calculations that require correction~\cite{komsa2014charged, komsa2018erratum, freysoldt2018first}.

Several computational and experimental studies investigated neutral point defects in phosphorene~\cite{cai2016highly, babar2016transition, ziletti2015oxygen, kiraly2017probing}. However, only a few address the charged defects in phosphorene~\cite{zhan2019interplay, guo2015vacancy, gaberle2018structure} even though in field-effect transistors, these native point defects can become charged and reduce the charge carrier mobility~\cite{liu2014phosphorene}. The previous DFT studies disagree on the charge transition levels (CTLs) of the phosphorene vacancy. Guo and Robertson~\cite{guo2015vacancy} reported a [+1/$-1$] CTL 0.24~eV above the valence band maximum (VBM), while Gaberle and Shluger~\cite{gaberle2018structure} predicted a [0/$-1$] CTL at 0.55~eV. In the previous studies the effect of defect configuration on the electronic structure has not been thoroughly investigated. Therefore, there is a need to conduct a comprehensive study of the effect of the structure and apply recently developed correction schemes~\cite{freysoldt2018first} to accurately calculate and validate the formation energy
% of charged vacancies
and charge transition levels of vacancies in phosphorene.

The prevalent approach~\cite{cohen1975self} for defect calculations embeds the defect into periodic supercells of increasing size to reduce the defect-defect interactions and converge the defect properties to the dilute limit~\cite{van2004first, shim2005density, leslie1985energy}. However, additional care must be taken when performing such calculations for charged defects due to spurious electrostatic interactions between the defect and its periodic images and the homogeneous compensating background charge.

In this work, we calculate the vacancy formation energy in phosphorene considering the $q = -1, 0, +1$ charge states. We employ the Freysoldt-Neugebauer correction method~\cite{freysoldt2018first} to account for the spurious electrostatic interaction between the defect charge and its images and the compensating background charge. We show that this method enables calculations of defect formation energies to within 0.1~eV accuracy in moderately sized simulation cells of 96 and 140 atoms. We discuss the electronic character of the defect state in terms of the relevant orbitals and the projected density of states. We predict a vacancy formation energy of 1.7~eV for the neutral vacancy and a [0,$-1$] CTL at 1.04~eV above the valence band maximum (VBM). This deep acceptor level exhibits a strong electron-lattice coupling with a large Stoke shift of 0.3~eV that could be measured by photoluminescence. The results show that intrinsic vacancy defects in phosphorene lead to deep acceptor levels that can compensate n-type dopants, trap charges, and scatter electrons, thereby reducing the carrier concentration and mobility.

\section {Method}
\subsection{Density Functional Theory}

We perform DFT~\cite{hohenberg1964inhomogeneous, kohn1965self} calculations to characterize the stability, charge transition levels, and density of states of various structures of the vacancy in phosphorene. We use the plane-wave code VASP~\cite{kresse1993ab, kresse1994ab, kresse1996efficiency, kresse1996efficient} with the projector augmented wave method~\cite{blochl1994projector}. For the exchange-correlation functional, we employ the Perdew-Burke-Ernzerhof (PBE)~\cite{perdew1996generalized} generalized gradient approximation (GGA) and the strongly constrained and appropriately normed (SCAN) meta-GGA approximation~\cite{Sun2015SCAN}. A plane-wave basis cutoff energy of 520~eV and a $k$-point density of 1000 per reciprocal atom for the Brillouin-zone integration with a Monkhorst-Pack mesh~\cite{monkhorst1976special} ensure energy convergence to within 1~meV/atom. We perform spin-polarized calculations and relax the atomic structures until the energy difference between subsequent interactions was smaller than 0.001\,meV for the simulation cell.

First, we relax both lattice vectors and ionic positions of the pristine rectangular four-atom unit cell of single-layer phosphorene obtained from MaterialsWeb.org~\cite{michaelpaper} to determine the equilibrium lattice parameters, shown in Tab.~\ref{tab:table1}. Then, from the relaxed unit cell, we create supercells of size $4\times3$, $6\times4$, and $5\times7$ with 48, 96, and 140 atoms, respectively. For the defect calculations, we remove one phosphorus atom per supercell, fix the lattice parameters to the equilibrium lattice constant of the pristine material, and relax only the ionic positions. This constraint enables the calculation of the elastic dipole of the defect, i.e., the derivative of the defect energy with respect to strain, and hence the calculation of the formation energy for small changes in the lattice constant if desired. To determine the energy convergence with respect to vacuum spacing, we investigate interlayer spacings of 10, 20, and 30~\AA. 

\begin{table}
\caption{\label{tab:table1}Calculated lattice parameters $a$ and $b$, fundamental bandgap $E_g$, and averaged permittivity, $\epsilon_r$, of phosphorene, calculated with the PBE and SCAN functionals compared to literature and experimental values.}
\begin{ruledtabular}
\begin{tabular}{lcccc}
& $a$ (\AA) & $b$ (\AA) & $E_g$ (eV) & $\epsilon_r$ \\
\colrule
PBE & 3.29 & 4.62 & 0.91 & 15.44 \\
SCAN & 3.28 & 4.60 & 1.28 & 12.50\\
HSE06 & & & 1.6 \\
\colrule
PBE~\cite{wang2015native} & 3.30 & 4.61 &  0.91 \\
HSE06~\cite{wang2015native} & 3.30 & 4.50 &  1.56 \\
$G_0W_0$~\cite{Ferreira2017} & & & 2.06 \\
Exp.~\cite{Liang2014, wang2015highly} & & & 2, 2.2 \\
\end{tabular}
\end{ruledtabular}
\end{table}

Table~\ref{tab:table1} shows the lattice parameters, fundamental bandgap, and permittivity (static dielectric constant) for phosphorene calculated using the PBE and SCAN functionals. The lattice parameters match well for both functionals. However, the calculated band gaps underestimate the experimental fundamental bandgap by about  1~eV, typical for semi-local exchange-correlation functionals. The meta-GGA functional SCAN somewhat improves the bandgap, and the hybrid functional HSE06~\cite{heyd2003hybrid} yields a value closest to experiment. The experimental optical bandgap of phosphorene is 1.3~eV due to a strong exciton binding energy of 0.9~eV~\cite{wang2015highly}. To accurately reproduce the fundamental and optical gap of phosphorene requires computationally demanding many-body $G_0W_0$ and Bethe-Salpeter calculations~\cite{Ferreira2017}, beyond the scope of this work.

\subsection{Defect Formation Energy and Charge Transition Level}

The formation energies of point defects in different charge states in semiconductors determine their charge transition levels (CTLs) and equilibrium concentrations as a function of the Fermi level~\cite{zhang1991chemical, van2004first}. The defect formation energy, $E_\mathrm{f}[X^q]$, of a defect $X$ with charge $q$ is given by
\begin{equation}
\begin{split}
E_\mathrm{f}[X^q] = & E_{\rm{tot}}[X^q]-E_{\rm{tot}}[\textrm{host}]- \\
& \sum_i n_i\mu_i + q \left ( \epsilon_F+\epsilon_v \right ) + E_\mathrm{corr},
%& \sum_i n_i\mu_i + q \epsilon_F + E_\mathrm{corr},
\end{split}
\end{equation}
%{\bf [RGH: You: If we reference the defect formation energy to the VBM and measure the Fermi energy relative to the VBM, then we do not need to add here the energy of the VBM again. Otherwise, in Eq. (3), we would need to have $\epsilon_F = \epsilon_v$. In other words, our energy reference for Eqs. (1) and (3) is $\epsilon_v = 0$.]}
where $E_{\textrm{tot}}[X^q]$ and $E_{\textrm{tot}}[\textrm{host}]$ are the total energies of the supercell containing the defect $X$ and the pristine host structure, respectively, $n_i$ denotes the number of atoms of type $i$ added or removed by the creation of the defect, and $\mu_i$ denotes their chemical potentials. For the single-component material phosphorene, the chemical potential $\mu_P$ is the energy per atom of pristine phosphorene. For the charged defects, the Fermi level, $\epsilon_F$, represents the energy of the electron reservoir in the material, and we reference $\epsilon_F$ to the VBM of the host material, $\epsilon_v$. The term $E_\mathrm{corr}$ comprises the correction for the spurious electrostatic interactions between the charged defect and its periodic images and the homogeneous compensating background charge~\cite{leslie1985energy, makov1995periodic, rozzi2006exact, freysoldt2009fully}.

In 2D materials, the compensating charge background generates a quadratic potential across the vacuum region, which causes an unphysical linear divergence of the monolayer energy with increasing vacuum spacing~\cite{komsa2014charged, komsa2018erratum, freysoldt2018first}. We utilize the Freysoldt-Neugebauer correction, which is applied in postprocessing and neither requires additional defect calculations nor relies on fitted parameters. The correction energy is obtained from a Gaussian approximation for the local defect charge density and aligns the long-range potential calculated for the defect potential with the model potential~\cite{freysoldt2009fully, freysoldt2011electrostatic, freysoldt2018first}. The method requires as input the electrostatic potential of the host and defect simulation cell and the permittivity. We calculate the permittivity using DFT, assuming that the monolayer is dielectrically isotropic, i.e., the in-plane ($\epsilon_\parallel$) and out-of-plane components ($\epsilon_\bot$) are equal. The computed permittivity tensor ($\epsilon_{\textrm{sc}}$) includes contributions from both the slab and the vacuum spacing in the unit cell, with the in-plane components acting in parallel and the out-of-plane component acting in series, resulting in the averaged permittivity for the monolayer of~\cite{freysoldt2008screening, Tan2020a}
\begin{equation}
\epsilon^{\textrm{slab}}=\frac{\epsilon^{\textrm{sc}}_\parallel - 1}{1-(\epsilon^{\textrm{sc}}_\bot)^{-1}}.
\end{equation}

Next, we calculate the thermodynamic charge transition level [$q/q'$] which is defined as the Fermi energy at which the defect formation energy of the charge state $q$ ($E_{\textrm{f}}(X^q)$) equals that of the charge state $q'$ ([$E_{\textrm{f}}(X^{q'})$]),
\begin{equation}
\label{eq:CTL}
[q/q']=\frac{E_\mathrm{f}(X^q;\epsilon_F=0)-E_\mathrm{f}(X^{q'};\epsilon_F=0)}{q'-q},
\end{equation}
where, $E_\mathrm{f}(X^q,\epsilon_F=0)$ is the defect formation energy of $X^q$ at the VBM.
The thermodynamic or adiabatic transition levels correspond to sufficiently slow processes in which the system fully relaxes to its new ground state. Hence the formation energies in Eq.~\eqref{eq:CTL} are evaluated for the relaxed charged defects. For instantaneous transitions occurring on time scales much shorter than lattice relaxations, such as in optical excitations, the atomic configuration is frozen. These optical transition levels are obtained from the formation energies for the different charge states with the same atomic configuration.\cite{freysoldt2014first}
In general, specific finite-size corrections for optical (vertical) transitions from charged initial states are required to account for the frozen-in screening response of the ions~\cite{Gake2020, falletta2020finite}. In phosphorene, however, the screening is purely electronic since the Born effective charges vanish for symmetry reasons. The corrections therefore reduce to the difference of the standard charge corrections according to the initial and final charge state, respectively.

\section {Results and Discussion}

\subsection{Defect structure}
 
The relaxation of the neutral vacancy defect leads to two different configurations depending on the initial symmetry. Starting from the pristine phosphorene supercell with one atom removed and preserving the mirror symmetry leads to the 5-5-6-6 ring structure~\cite{hu2015defects, babar2016transition}. However, when breaking the symmetry by perturbing a neighboring atom to the vacancy, the structure relaxes to the lower energy 9-5 ring configuration~\cite{babar2016transition} shown in Fig.~\ref{fig:phosphorene}(a). The observed distortion for the neutral vacancy in phosphorene is similar to the reconstruction surrounding the vacancy in graphene, which leads to the formation of a new bond saturating two of the three dangling bonds with a single dangling bond remaining for geometric reasons~\cite{robertson2013structural,banhart2011structural} and giving rise to a five- and nine-membered ring structure. However, Guo and Robertson~\cite{guo2015vacancy} reported that the negative vacancy in phosphorene does not display the 9-5 ring structure of the neutral and the positive vacancy. We did not observe such discrepancy and find that the 9-5 ring structure is the lowest in energy. We note that if the symmetry is not broken during the relaxation process, the structure relaxes to the higher energy metastable 5-5-6-6 structure~\cite{guo2015vacancy}. The discrepancy may be due to the use of different functional as they use screened exchange (sX) hybrid density functional.~\cite{clark2010screened}

The calculated vacancy defect formation energy for the 5-5-6-6 structure is 1.95~eV, whereas the 9-5 structure has a lower energy of 1.59~eV (PBE). The calculated defect formation energies for both structures match those reported in the literature~\cite{hu2015defects, huang2018optically, babar2016transition}. 
The new bonds which form the 9-5 rings have bond length of 2.33~\AA,  which is larger than the bond length in pristine phosphorene of 2.24~\AA. The relaxed structures of the neutral and +1 charged P vacancy are essentially identical. However, the ring structure around the negatively charged defect is further distorted; the 4 bonds on the left side of the ring shrink while the ones on the right increase in length. The changes in bond lengths are less than 0.1~\AA, hence, the 9-5 ring structure is maintained.

\begin{table}
\caption{\label{tab:table2}Calculated defect formation energy for neutral vacancy defect with the two ring structure. The ring structure depends upon the broken and the unbroken symmetry}
\begin{ruledtabular}
\begin{tabular}{lcc}
Structure & PBE & SCAN  \\
\colrule
9-5 & 1.64~eV & 1.72~eV \\
5-5-6-6 & 1.95~eV & 2.28~eV\\
\end{tabular}
\end{ruledtabular}
\end{table}

Next, we calculate the formation energies of the charged vacancy defects and the charge transition levels and characterize the electronic structure of the lower energy 9-5 ring structure.

\subsection{Energetics}

%\begin{figure}
%    \includegraphics[width=\columnwidth]{figures/convergence_vacancy_defect_nicecolor.pdf}
    
%    \caption{\label{fig:EnergyCorrection}(Color online) Uncorrected (unfilled symbols) and corrected (filled symbols) vacancy formation energies for different charged states, cell sizes, and vacuum spacings, calculated using the PBE functional at the Fermi level corresponding to the valence band maximum. For each supercell, the uncorrected formation energies increase for increasing vacuum spacing from 10 to 30~\AA.} 
%\end {figure}\

%Figure~\ref{fig:EnergyCorrection} compares the uncorrected and corrected defect formation energies for the vacancy in 96 and 140-atom simulation cells with interlayer vacuum spacings of 10, 20, and 30~\AA.
%The calculations in the 48-atom cells did not converge {\bf[RGH: Can we be more specific by stating if the structure relaxed to a different configuration or something else happened or we could leave this sentence off?]}, indicating that defect calculations for phosphorene require larger cell sizes.

We compare the corrected defect formation energies for the vacancy in 96 and 140-atom simulation cells with interlayer vacuum spacings of 10, 20, and 30~\AA. For the neutral defect in the 96 and 140-atom cells, the formation energies changes by less than 10~meV when the vacuum spacing increases from 20 to 30~\AA. For the charged cases, the energy correction removes the linear divergence of the uncorrected formation energy with increasing vacuum spacing. The corrected formation energies of the charged vacancy in the 96 and 140-atom cells differ by 70 and 5~meV, respectively, for vacuum spacings of 20 and 30~\AA.
%For the neutral defect in the 96 and 140-atom supercells, the defect formation energies changed by less than 10~meV when increasing the vacuum spacing from 20 to 30~\AA. For the charged cases, the energy correction removes the linear divergence of the uncorrected formation energy with increasing vacuum spacing. For the 140 atoms supercell, the difference in energy between vacuum spacing is less than 5~meV after the correction scheme is applied. As 
%We also notice that the added correction term is higher (0.4~eV) for 10 vacuum spacing than the 30 vacuum spacing (0.1~eV) indicating that the smaller vacuum spacing has larger electrostatic interaction.]}  {\bf [RGH: As Anne Marie points out, the FN method corrects the electrostatic interlayer interactions. After correction, which vacuum spacings deviate from the others? It appears that the larger vacuum spacing of 30\AA\ deviates the most. The largest vacuum spacing has the highest unphysical electric field in the vacuum region that could slightly change the structure and lead to small energy differences You could check how much the P-P distances differ for the 30\AA\ case.

The corrected formation energies for the neutral and negatively charged vacancy show a similar convergence with respect to supercell size, indicating that the energy correction removes the in-plane electrostatic interactions between the charged defects. The formation energy of the neutral vacancy changed from 1.64~eV for the 96 atom cell to 1.59~eV for the 140 atom cell. The formation energy for the positively charged vacancy exhibited a stronger dependence on the in-plane cell size due to the defect state being below the VBM, resulting in the excess charge being delocalized in the 2D material, which is not accounted for by the correction scheme. Overall, we estimate that the defect formation energies are converged to within about 0.1~eV for the supercells of 96 atoms and vacuum spacings of 20~\AA\ and use this supercells for the characterization of the vacancy.

\begin{figure}
    \includegraphics[width=\columnwidth]{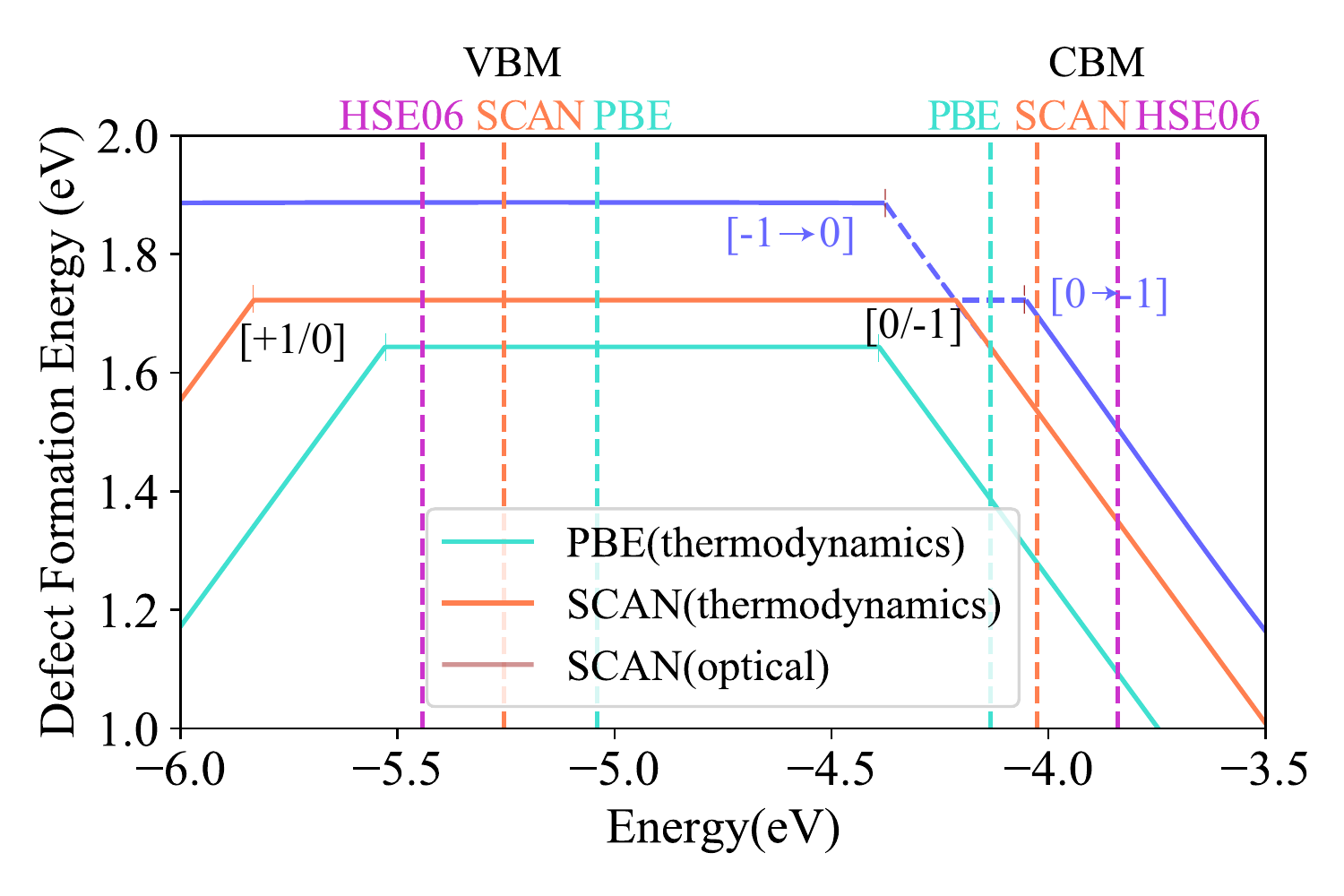}
    \caption{\label{fig:CTLs}(Color online) Defect formation energies of the charged vacancy defects in phosphorene as a function of Fermi level calculated with the PBE and SCAN functional. The resulting thermodynamic and optical charge transition levels are compared with the valence and conduction band edges from PBE, SCAN, and HSE06 using the vacuum level as a common reference. The vacancy only shows a [0/$-1$] charge transition inside the bandgap, close to the conduction band minimum. The differences in energy between the thermodynamic  and the optical transition levels corresponding to absorption $0 \rightarrow -1$ and emission $-1 \rightarrow 0$ indicates large Franck-Condon shifts for the deep acceptor level of the vacancy.} 
\end {figure}

Fig.~\ref{fig:CTLs} shows the formation energies of the charged vacancy obtained with the PBE and SCAN functional and compares the resulting thermodynamic and optical charge transition levels to the band edges obtained with the PBE, SCAN, and HSE06 functional using vacuum level as the common reference.
%The colored lines represent the formation energies of the vacancy in the different charge states, with the slopes of each line segment corresponding to the defect charge.
%The turquoise and orange colored lines represent defect formation energies calculated with PBE and SCAN functionals respectively, in which the defect atomic structure is fully relaxed to the equilibrium configuration. Therefore, the two lines show the formation energy of the vacancy in its most stable charge state at a given Fermi level, and the kinks on the lines indicate the charge transition levels (CTLs) where the vacancy defect is predicted to undergo a thermodynamic charge transition.
%For PBE, the +1 vacancy is only stable at energies below the VBM, the neutral vacancy is stable in the energy range from the VBM to 0.65~eV above the VBM in the band gap, while the $-1$ vacancy is stable at energies above that. 
The vacancy exhibits a [0/$-1$] charge transition inside the bandgap at 0.65 and 1.04~eV above the VBM for the PBE and SCAN functional, respectively.  The [0/$-1$] charge transition occurs close to the conduction band minimum (CBM), indicating that the vacancy in phosphorene acts as a deep acceptor. The [+1/0] charge transition falls well inside the valence band region, indicating that positively charged vacancies do not occur in monolayer phosphorene. For the 6-6-5-5 ring structure, which is not the ground state, the charge transition levels are outside the band gap region. 
\begin{figure*}[t]
    \center
    \includegraphics[width=\textwidth]{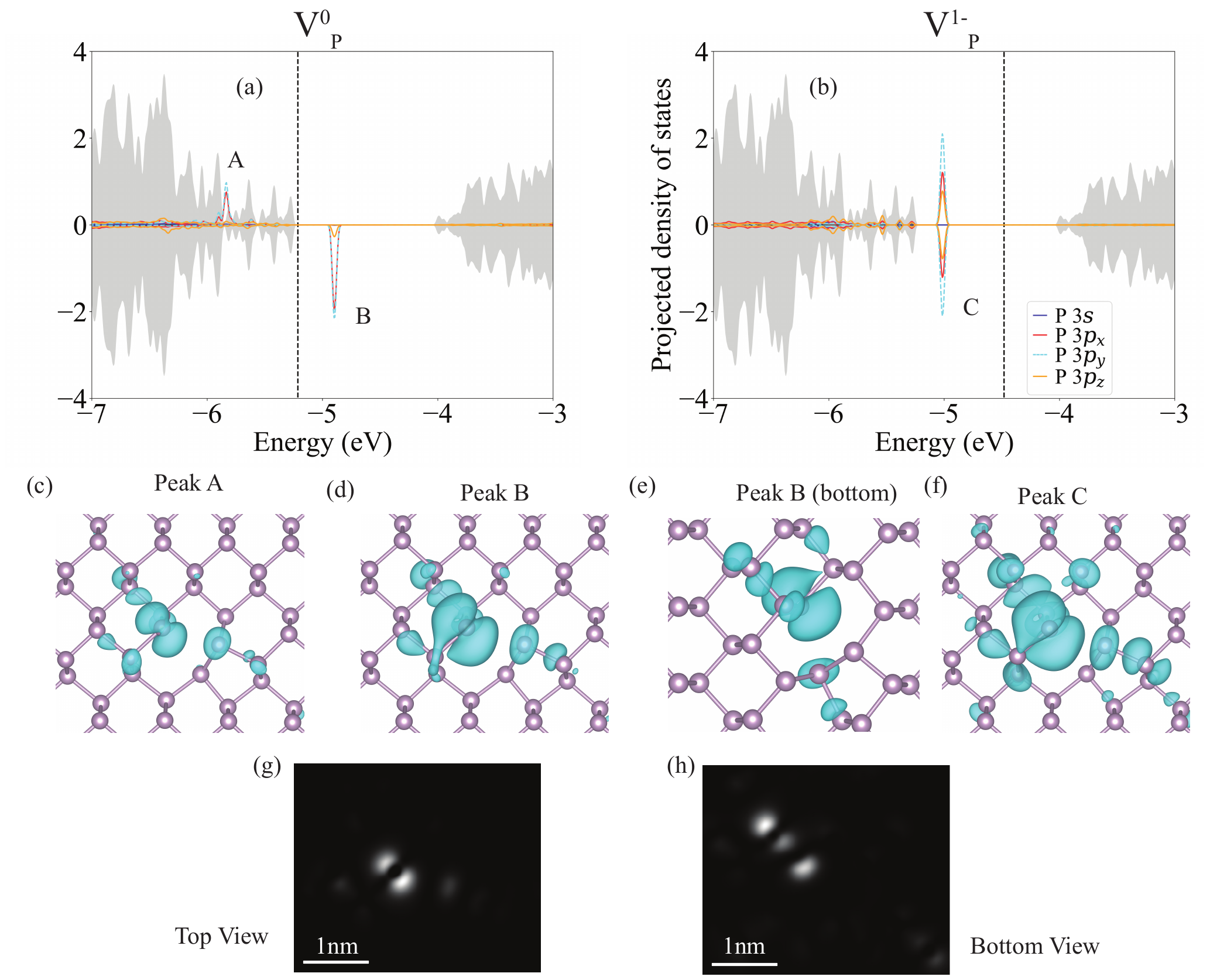}
    \caption{\label{fig:dos}(Color online) DOS for (a) the neutral and (b) the negatively charged vacancy projected onto the $s$ and $p$ orbitals of the atoms nearest to the vacancy defect calculated with the SCAN functional. The grey shaded regions indicate the DOS of pristine phosphorene. The energies are plotted with respect to the vacuum level. The dashed line indicate the Fermi level. (c)--(f) Isosurfaces of the charge density for the defect levels labeled A, B, and C (top and bottom view of the slab). (g,h) The top and bottom view of the simulated STM image for peak B in the density of states. The STM images are nearly identical for the defect level in the bandgap of the neutral and negatively charged vacancy.} 
\end {figure*}
Optical transitions occur on time scales too fast for structural relaxations. Therefore, the energy difference between the neutral and negatively charged vacancy without relaxations provides an estimate of the optical CTLs for absorption [$0 \rightarrow -1$] of $-4.1$~eV  and emission [$-1 \rightarrow 0$] of $-4.4$~eV relative to the vacuum level. The large observed shift in the optical CTL relative to the thermodynamic CTL in Fig.~\ref{fig:CTLs} indicates sizable Franck-Condon shifts and a combined Stokes shift of 0.3~eV for the phosphorene vacancy that could be measured by photoluminescence. 

The intrinsic phosphorene vacancy defect could be neutral or negatively charged depending upon the Fermi level position. This could lead to two possible excitation and emission processes. When the defect is negatively charged, such as in $n$-type doped phosphorene, an electron could be excited from the filled defect state to the CBM absorbing a photon of 0.38/0.57~eV energy (SCAN/HSE06), and the emission would occur when an electron jumps from the CBM to the defect state with the emitted photon energy of 0.09/0.28~eV (SCAN/HSE06). Similarly, when the defect is charge neutral, such as in undoped or $p$-doped phosphorene, an electron could be excited from the VBM to the empty defect state, creating a hole in the VBM and absorbing 0.84/1.03~eV energy (SCAN/HSE06), and the emission would occur when an electron moves from the defect state to fill the VBM hole emitting a photon of 1.13/1.32~eV energy (SCAN/HSE06). The SCAN functional predicts that this entire process takes place in the infrared and the microwave regime depending upon the nature of the defect charge. The more accurate band edges from the HSE06 functional indicate that the emission and absorption occur in the infrared region.

With the Stoke's shift in the infrared region, phosphorene could have applications in biosensing and solar concentrators. Doped graphene quantum dots based fluorescence biosensor have shown remarkable biosensing ability that have potential applications in biomedical applications~\cite{kalkal2020biofunctionalized}.
Also, tandem solar concentrators with two different absorption onsets, one in the near infrared region and the other in the visible region have demonstrated improved power conversion efficiency for collecting sunlight~\cite{wu2018tandem}.

\subsection{Density of States}

%, charge isosurfaces for Peaks A-C [c-e] and the simulated STM image[f]. Peaks A -- C give almost identical STM images, hence only one image is shown.
%The projected density of states corresponding to the single P vacancy in different charge states in single layer phosphorene Fig.~\ref{fig:dos} shows that the defect level with predominantly $p_x$ and $p_y$ orbital character is localized between the band edges. The density of states is projected onto the $s$- and $p$-orbitals of the P atom nearest to the P vacancy. In the neutral defect there are two peaks depicting the defect level. The defect level indicated by Peak A in Fig.~\ref{fig:dos}(a) in the valence band is filled while Peak B in the same figure within the band gap is empty. This empty state is an deep acceptor level as the energy required to excite an electron from the VBM to fill the state -- indicated by its distance away from the VBM -- is greater than the thermal energy. When the system is negatively charged, the extra electron occupies the previously empty state which has predominantly $p_x$ and $p_y$ character as seen in Peak C in Fig.~\ref{fig:dos}(b). The states that are filled below the VBM ($-6.5$ and $-7.0$~eV) are shifted to the top of the valence band and form completely filled degenerate state.

To characterize the electronic structure of the vacancy, we calculate the electronic density of state (DOS) and charge density. Fig.~\ref{fig:dos}(a,b) shows the DOS projected onto the $s$ and $p$ orbitals of the P atoms neighboring the vacancy for the negatively charged and neutral defect, illustrating that the defect state predominantly exhibits $p_x$ and $p_y$ orbital character. For the neutral vacancy, the singly occupied defect level is spin-polarized. The occupied state (A) occurs inside the valence band while the empty state (B) remains in the bandgap. For the negative vacancy, the defect level (C) becomes doubly occupied and is inside the bandgap. Fig.~\ref{fig:dos}(c-e) illustrates the charge density of the defect orbitals corresponding to peaks A, B, and C. The defect charge density is strongly localized on the $p$ states of the P atoms surrounding the vacancy and extends in the $x$ and $y$ directions, validating the DOS results. The DOS is shown with respect to the vacuum level. Calculating the vacuum level for the neutral case is straightforward; however, care must be taken when calculating the vacuum energy for the charged cases. First we obtain the vacuum level in the pristine supercell. Then we subtract the contribution from the electrostatic artifact obtained from FNV correction scheme to remove the unphysical quadratic potential and account for the alignment term due to the background charges.
%and not in the $z$ direction, indicating that the electrons near the defect are in the $p_x$ and $p_y$ orbitals, validating the results from the density of states discussed earlier.

Comparing defect orbitals obtained from DFT with scanning tunneling microscopy (STM) $dI/dV$ images can help verify a defect's type and nature~\cite{Kozhakhmetov2020}. Fig.~\ref{fig:dos}(f,g) shows the simulated STM images from the top and the bottom of the monolayer obtained from the projected charge density between 0.25-0.4~eV above the VBM that corresponds to the peak B in the density of states. The top view shows a characteristic asymmetry in the simulated STM image for the defect state inside the bandgap of the neutral vacancy with the 5-9 ring structure that can help experimental validation. The STM images are nearly identical for the negatively charged vacancy. The image shows that the vacancy defect state is dominated by a $p$ orbital localized on the neighboring P atom, similar to its charge density.

Importantly, the energy of the peaks for the defect states in the DOS differ from the CTL's since the DOS and band structure illustrate the one-electron Kohn-Sham energies while the true CTL's correspond to total energy differences of different charge states that include the many-electron contributions to the energy from exchange and correlation and the charge correction. Therefore, the DOS cannot accurately predict the position of CTL's.

The choice of exchange-correlation functionals affects the position of the band edges and CTL's. %Thus, it is important to compare different functionals as we have performed in this work to investigate defect properties for reliable and accurate results.
In this study, PBE and SCAN predict similar defect formation energies leading to the qualitatively same conclusions. Similarly, we do not expect the usually more accurate HSE06 functional to change the conclusion of our results. In our previous study of the vacancy in 2D MoS$_2$, the HSE06 functional predicts a larger bandgap than SCAN but retains the [0/$-1$] CTL at the same position relative to the vacuum level.\cite{tan2021complexes} If assuming a similar behavior in phosphorene, the [0/$-1$] CTL would occur about 1.2~eV above the VBM. 

We compare our structure and charge transition level with two previous studies. Guo and Roberson~\cite{guo2015vacancy} reported a [+1/$-1$] CTL within the bandgap. However, the applied charge correction~\cite{lany2008assessment, persson2005n} did not account for the varying dielectric profile and large vacuum spacing specific to 2D materials. Gaberle and Shluger~\cite{gaberle2018structure} reported a [0/$-1$] CTL at 0.55~eV above the VBM, which closely agrees with our result of 0.7/1.04~eV for PBE/SCAN. For charge correction,  they extrapolate the defect formation energy to the infinite supercell.~\cite{Guidon2009} The difference in the CTL energy may be due to the  choice PBE0-TC-LRC hybrid functional and the charge correction method. 

\section {Conclusions}
We computed the relaxed strucure and defect formation energies for charged single vacancies in monolayer phosphorene, utilizing the charge correction scheme developed by Freysoldt and Neugebauer to restore the appropriate electrostatic boundary conditions. We find that the symmetry broken 9-5 ring configuration is the ground state structure for the neutral and charged vacancies. The neutral vacancy has a formation energy of 1.7~eV and undergoes a charge transition to the negative state at a Fermi level of 1.04~eV above the valence band maximum within the bandgap. Hence, phosphorene's single vacancy acts as a deep acceptor that could passivate dopants in $n$-type phosphorene and reduce the carrier concentration and mobility. We predict that the defect level in the bandgap exhibits $p$-orbital character that can be revealed by scanning tunneling microscopy. The optical charge transition levels display sizeable Frank-Condon shifts and a large Stokes shift of more than 0.3~eV, providing possible opportunities for biosensing and photovoltaics applications.

\begin{acknowledgments}
This work was supported by the National Science Foundation under Grant Nos.\ DMR-1748464 and OAC-1740251 and the 2DCC-MIP under Grant No.\ DMR-1539916. Computational resources were provided by the University of Florida Research Computing Center. Part of the research was performed while the authors visited the Institute for Pure and Applied Mathematics (IPAM), which is supported by the National Science Foundation (Grant No.\ DMS-1440415).
\end{acknowledgments}

%\bibliography{references}
%merlin.mbs aipnum4-1.bst 2010-07-25 4.21a (PWD, AO, DPC) hacked
%Control: key (0)
%Control: author (8) initials jnrlst
%Control: editor formatted (1) identically to author
%Control: production of article title (-1) disabled
%Control: page (0) single
%Control: year (1) truncated
%Control: production of eprint (0) enabled
%

\end{document}